\documentclass{article}
\usepackage{spconf,amsmath,graphicx}
\usepackage{times}
\usepackage{epsfig}
\usepackage{graphicx}
\usepackage{subcaption}
\usepackage[dvipsnames]{xcolor}
\usepackage{amsmath}
\usepackage{amssymb}
\newtheorem{theorem}{Theorem} 
\usepackage{siunitx}
\usepackage{float}
\usepackage{subcaption}

\let\vec\mathbf

\title{Sensor Networks TDOA Self-Calibration: 2D Complexity Analysis and Solutions}
\name{Luca Ferranti $^{a,c}$, Kalle Åström $^b$, Magnus Oskarsson $^b$, Jani Boutellier $^a$, Juho Kannala $^c$}
\address{$^a$University of Vaasa, Vaasa, Finland, $^b$Lund University, Lund, Sweden\\ $^c$Aalto University, Espoo, Finland}
\begin{document}
%
\maketitle
\begin{abstract}
Given a network of receivers and transmitters, the process of determining their positions from measured pseudoranges is known as network self-calibration. In this paper we consider 2D networks with synchronized receivers but unsynchronized transmitters and the corresponding calibration techniques, known as Time-Difference-Of-Arrival (TDOA) techniques. Despite previous work, TDOA self-calibration is computationally challenging. Iterative algorithms are very sensitive to the initialization, causing convergence issues. In this paper, we present a novel approach, which gives an algebraic solution to two previously unsolved scenarios. We also demonstrate that our solvers produce an excellent initial value for non-linear optimisation algorithms, leading to a full pipeline robust to noise.
\end{abstract}
\begin{keywords}
Time Difference of Arrival, Sensor Networks Calibration, Minimal Problems
\end{keywords}

\section{Introduction}
\label{ch:intro}
Wireless Sensor Networks have been widely studied \cite{yick2008} and have been successfully applied to several domains, such as positioning \cite{savarese2002}, mapping \cite{slat}, microphone array calibration \cite{plinge2016acoustic} and beamforming \cite{ochiai2005}.
In order to be properly used, the network must first be calibrated, i.e. the positions of its nodes need to be determined. This can be done e.g. using transmitters at a known position and trilaterating the network nodes. However, several applications require simultaneous localization of both receivers and transmitters \cite{muller2011}. This is known as \textit{network self-calibration} \cite{miluzzo2008, yick2008} and it is the main theme of this paper.

Let us consider a network with $m$ receivers and $n$ transmitters, shortly denoted as $mr/ns$. From the measurements $mn$ equations will be available to solve the self-calibration problem. Now, we consider two different scenarios:
\begin{itemize}
    \item \textbf{Synchronized RXs and TXs}: In this situation, the time instants at which the signal is transmitted and measured are known and hence the distances between RX and TX are measured. For each pair of receiver-transmitter, the following polynomial equation is obtained
    \begin{equation}
    f_{ij}^2=d_{ij}^2=\Vert\vec{r}_i-\vec{s}_j\Vert^2, \label{eq:toa1}
    \end{equation}
    where $f_{ij}$ is the measured range and $d_{ij}$ is the distance between the receiver located at $\vec{r}_i$ and the transmitter located at $\vec{s}_j$. 
    In total we will have $K(m+n)-G$ degrees of freedom (DoF), where $K$ denotes the spatial dimension ($K=2,3$) and $G$ is the \textit{Gauge freedom} ($G=3$ in 2D and $G=6$ in 3D). Since we measure only distances, the positions can be recovered only up to a Euclidean transformation. Practically, this means that the coordinate systems in which we solve the coordinates can be chosen freely, reducing the degrees of freedom. This formulation is known as \textit{Time-Of-Arrival} (TOA).
    \item \textbf{Synchronized RXs and unsynchronized TXs}: In this situation, all receivers will measure the time of arrival in the same clock frame. For each pair of transmitter-receiver, the following polynomial equation is obtained
    \begin{equation}
        (f_{ij}-o_j)^2=d_{ij}^2=\Vert\vec{r}_i-\vec{s}_j\Vert^2. \label{eq:tdoa}
    \end{equation}
    As equation \eqref{eq:tdoa} shows, each transmitter introduces an extra unknown $o_j$, the time offset between the transmitter local clock and the receivers clock. The total number of degrees of freedom will thus be $K(m+n)+n-G$. This formulation is known as \textit{Time-Difference-Of-Arrival} (TDOA).
\end{itemize}
The above mentioned equations could be solved by numerical iterative methods \cite{priyantha2003, biswas2004}. These approaches, however, can suffer from several issues such as getting stuck in local minima, slow convergence and sensitivity to outliers. It has been shown that algebraic non-iterative approaches can achieve higher accuracies \cite{kuang2013, kuang2013tdoa, batstone2016, crocco2012, larsson2017cvpr}. Furthermore, using the algebraic solution as initial value for iterative methods allows a faster and more accurate convergence.

This paper focuses on TDOA, particularly to the 2D case ($K=2,~G=3$). The total number of DoF will thus be $2m+3n-3$. The configurations for which the number of DoF equals the number of equations are referred to as \textit{minimal configurations} and the problems of network self-calibration with the minimum amount of receivers and transmitters are called \textit{minimal problems}. The minimal configurations for 2D TDOA are 6r/3s and 4r/5s. Previous work developed algebraic solvers for non-minimal cases, such as 7r/6s, 5r/6s \cite{kuang2013tdoa} and 8r/4s \cite{pollefeys2008}, however the minimal problems cannot be solved algebraically yet.

\begin{table}[t]
    \centering
    \caption{Different TDOA configurations. X: solved in \cite{kuang2013tdoa}. M: minimal. O: solved in this paper. *: reducible to a solved configuration. -: Unsolved. u: underdetermined}
    \label{tab:tdoa}
    \begin{tabular}{ccccc}
        \hline\hline
         $m/n$&3&4&5&6  \\\hline
         4&u&u&M&-\\
         5&u&-&-&X\\
         6&M&O&*&*\\
         7&-&X&*&*\\
         8&-&*&*&*\\
         9&O&*&*&*\\
    \end{tabular}
\end{table}

In this paper, we fill the gap towards minimal problems, proposing a new approach, able to solve two previously unsolved configurations: $9r/3s$ and $6r/4s$, as summarized in Table \ref{tab:tdoa}. Opposed to previous methods, where TDOA was tackled by first solving for the offsets and then solving the remaining TOA  problem, our approach combines TOA and TDOA ideas and jointly solves both offsets and positions, reducing the overall computation load of the pipeline. Furthermore, we provide a quantitative estimate of the computational complexity of several unsolved configurations.

The paper is structured as follows: in Section II the current state of the art of TOA and TDOA solving techniques is reviewed. In Section III our proposed method is explained, and our solvers are benchmarked in Section IV. Finally, conclusions are drawn in Section V.

\section{Related Work}
\label{ch:relatedWork}
In these sections, we review the state-of-the-art techniques that have been used to algebraically solve TOA and TDOA so far.

\subsection{TOA}
From the distances $d_{ij}$ we can define the \textit{compaction matrix} \cite{kuang2013} $\vec{\tilde{D}}\in\mathbb{R}^{(m-1)\times(n-1)}$ such that
\begin{equation}
    [\vec{\tilde{D}}]_{ij} = d_{i+1,j+1}^2-d_{1,j+1}^2-d_{i+1,j}^2+d_{11}^2.
\end{equation}

With algebraic manipulation, it can be shown that the following factorization holds 
\begin{equation}
    \vec{\tilde{D}}=-2\vec{R}^T\vec{S}, \label{eq:toa_fac}
\end{equation}
where $\vec{R}_i=[\vec{r}_{i+1}-\vec{r}_1]$ for $i=1\ldots m-1$ and similarly $\vec{S}_j=[\vec{s}_{j+1}-\vec{s}_1]$ for $j=1\ldots n-1$. 

This factorization is not unique, suppose we have \textit{a} factorization $\tilde{\vec{D}}=\tilde{\vec{R}}\tilde{\vec{S}}$. Clearly, for each full-rank matrix $\vec{L}$, it holds $\tilde{\vec{D}}=\tilde{\vec{R}}^T\vec{L}^{-1}\vec{L}\tilde{\vec{S}}=\tilde{\vec{R}}^T\tilde{\vec{S}}$. After $\tilde{\vec{R}}$ and $\tilde{\vec{S}}$ have been computed using e.g. Singular Value Decomposition, the problem is reduced to determine the matrix $\vec{L}$ so that $\vec{R}=\vec{L}^{-T}\tilde{\vec{R}}$ and $\vec{S}=\vec{L}\tilde{\vec{S}}$. The receivers and transmitters can now be parametrized as follows
\begin{equation}
    \begin{split}
        &\vec{r}_1=\vec{0}\quad \vec{s}_1=\vec{L}\vec{b},\\
        &\vec{r}_i=\vec{L}^{-T}\tilde{\vec{R}}_{i-1},\quad i=2\ldots m,\\
        &\vec{s}_j=\vec{L}\left(-\frac{1}{2}\tilde{\vec{S}}_{j-1}+\vec{b}\right),\quad j=2\ldots n,
    \end{split} \label{eq:toa_upgrade}
\end{equation}
where $\vec{b}$ is a vector to be determined. Finally, defining $\vec{H}=\left(\vec{L}^T\vec{L}\right)^{-1}$, the following equations can be derived
\begin{equation}
    \begin{split}
        &\mathrm{(A)}\quad d_{11}^2=\vec{b}^T\vec{H}^{-1}\vec{b},\\
    &\mathrm{(B)}\quad d_{1j}^2-d_{11}^2=\frac{1}{4}\tilde{\vec{S}}^T_{j-1}\vec{H}^{-1}\tilde{\vec{S}}_{j-1}-\vec{b}^T\vec{H}^{-1}\tilde{\vec{S}}_{j-1},\\
    &\mathrm{(C)}\quad d_{i1}^2-d_{11}^2=\tilde{\vec{R}}^T_{i-1}\vec{H}\tilde{\vec{R}}_{i-1}-2\vec{b}^T\tilde{\vec{R}}_{i-1},
    \end{split}\label{eq:toa}
\end{equation}
with $i=2\ldots m$ and $j=2\ldots n$. Since the matrix $\vec{H}$ is symmetric, it can be parametrized in $3$ unknowns and hence $\vec{H}$ and $\vec{b}$ will depend on $5$ parameters in total. Given $m$ receivers and $n$ transmitters, one equation of type (A) with degree $3$, $n-1$ equations of type (B) with degree $2$ and $m-1$ equations of type (C) with degree $1$ are obtained. A fast polynomial solver can finally be derived from these equations using Gröbner basis \cite{kukelova08, larsson2017cvpr, stewenius-phd-2005}. This approach was exploited in \cite{kuang2013} to solve the TOA minimal problems.

\subsection{TDOA}
For TDOA, $\vec{\tilde{D}}$ depends on the offsets and hence it cannot be factorized numerically. As a consequence, the parametrization of the previous section cannot be used. However, from the dimensions of $\vec{R}$ and $\vec{S}$, we notice that $\text{rank}~\tilde{\vec{D}}=2$. If $m>3$ and $n>3$, this means that $\vec{\tilde{D}}$ is rank deficient and thus all $3\times 3$ subdeterminants must be equal to zero. Despite having in general $\binom{m-1}{3}\cdot\binom{n-1}{3}$ subdeterminants, the following theorem holds \cite{kuang2013tdoa}.
\begin{theorem}
Given a rank 2 matrix $\vec{A}\in\mathbb{R}^{(m\times n)}$, $m,n>2$ then $(m-2)(n-2)$ independent rank constraints can be obtained.
\end{theorem}
Since the compaction matrix is $\vec{\tilde{D}}\in\mathbb{R}^{(m-1)\times(n-1)}$, given $m$ receivers and $n$ transmitters, $(m-3)(n-3)$ independent constraints can be obtained. This approach was used to solve some TDOA configurations in \cite{kuang2013tdoa}, where the rank constraints were used to determine all the offsets, reducing the problem to TOA. It is good to notice, however, that rank constraints alone cannot be used to solve the \textit{minimal} cases of TDOA. For the minimal problem 6r/3s, the compaction matrix will already have two columns, and hence no rank constraints can be derived. For the other minimal case, 4r/5s, only two independent constraints can be obtained, which is not enough to solve for the five unknown offsets. \textit{In this paper we propose a different symbolic factorization of the compaction matrix, which allows to exploit the parametrizations \eqref{eq:toa_upgrade}-\eqref{eq:toa}, leading to new solvers for previously unsolved cases.}

\section{Proposed Method}
\label{ch:method}
  In this section we describe the numerical techniques used to solve the problems arising in TDOA. The core idea is to use the factorization in \eqref{eq:toa_fac} to produce new equations depending on both offsets and coordinates. Opposed to previous TDOA approaches, we aim at solving all unknowns in one step. First, we show how a trivial factorization can be obtained when only three transmitters are present. Next, we show how this can be generalized to more transmitters. In all our formulations, we fix the Gauge freedom by imposing $\vec{r}_1=\vec{0}$ and $\vec{r}_2=[r_{2x},0]^T$.

\subsection{Three transmitters}

If only three transmitters are available, no rank constraints can be imposed. However, it can be noticed that the following holds
\begin{equation}
    \tilde{\Vec{D}}= (\tilde{\Vec{D}}^T)^T\Vec{I}, \label{eq:TDOA_fac}
\end{equation}
where $\vec{I}$, is the identity matrix. Hence, we can formulate the equations in \eqref{eq:toa} imposing $\tilde{\Vec{R}}=\tilde{\Vec{D}}^T$ and $\tilde{\vec{S}}=\vec{I}$, obtaining $m+2$ equations in $8$ unknowns ($5$ for $\vec{H}$ and $\vec{b}$ and $3$ for the offsets). The minimal case is, as previously shown, 6r/3s. It is good to notice that now equations of type (C), despite being linear in $\vec{H}$ and $\vec{b}$, are overall of degree 3, as $\tilde{\vec{R}}$ depends on the unknown offsets.
For the subminimal cases ($m>6$), we have more constraints than unknown. This raises the question \textit{how should we pick the eight equations from the $m+2$ available?} First, let us introduce the notation $abc$ to denote the formulation using $a$ equations of type A, $b$ equations of type B and $c$ equations of type C. Now for each formulation we compute the \textit{standard monomial basis} $\mathcal{B}$ \cite{cox-little-etal-98} associated with its polynomial ideal. It was shown in previous work \cite{kukelova08, larsson2017cvpr}, that the size of the standard monomial measure can be used as a complexity measure for the problem. The smaller it is, the faster and more accurate the final solver will be. The results of the simulation are shown in Table \ref{tab:3rec}. 

 \begin{table}[tbh]
     \centering
     \caption{Size of the standard monomial basis for different formulations using 3 receivers.}
     \begin{tabular}{ccc||c}
          A&B&C&$|\mathcal{B}|$  \\\hline
          0&0&8&75\\
          0&1&7&116\\
          1&0&7&160\\
          1&1&6&198\\
          0&2&6&144\\ 
          1&2&5&181
     \end{tabular}
     \label{tab:3rec}
 \end{table}

As can be noticed from Table \ref{tab:3rec}, equations of type C lead to the lowest computational complexity and equations of type A to the highest computational complexity. Particularly, for the 9r/3s case, the problem can be solved using only equations of type C. The last row of the table corresponds to the minimal problem 6r/3s. \textit{Observing Table \ref{tab:tdoa}, this new solver allows to calibrate a network using as little as $3$ transmitters, whereas previous state-of-the-art solvers could not handle networks with only $3$ transmitters.} 

Summarizing the previous discussion, we obtain the following final numerical receipt for the new $9r/3s$ solver:
\begin{enumerate}
    \item Formulate the polynomial equations using the factorization \eqref{eq:TDOA_fac} and the equations of type (C) from \eqref{eq:toa}.
    \item Using the techniques described in \cite{kukelova08} and \cite{larsson2017cvpr}, generate a solver for the polynomial system.
    \item Compute the receivers and transmitters positions using \eqref{eq:toa_upgrade}
    \item Refine the solution of the previous step using non-linear optimisation (e.g. Levenberg-Marquardt algorithm).
\end{enumerate}

\subsection{More than three transmitters}
If we have more than three transmitters, then $\tilde{\vec{S}}$ will not be square and it cannot be directly replaced by the identity matrix. Thus, in order to use the factorization in \eqref{eq:TDOA_fac}, we need to discard some transmitters. Particularly, let us consider the previously unsolved case $6r/4s$, using the parametrization in \eqref{eq:toa}, we obtain $9$ unknowns in total, $5$ for $\vec{H}$ and $\vec{b}$ and $4$ for the offsets. Now our compaction matrix $\tilde{\vec{D}}$ has size $5\times3$. To solve this configuration, we consider the new compaction matrix $\hat{\vec{D}}$, obtained discarding the last column of $\tilde{\vec{D}}$. Using the same method described in the previous section, we obtain $1$ equation of type (A), $2$ equations of type (B) and $5$ equations of type (C), i.e. 8 equations in total. Furthermore, since the original compaction matrix $\tilde{\vec{D}}$ has to be rank $2$. Imposing all $3\times3$ minors to $0$, we obtain $10$ equations in the offsets, out of which only $3$ are independent by Theorem 1. Again, we have computed the standard monomial basis for different combinations of the above mentioned equations and determined that the most efficient solver was obtained including all $5$ equations of type (C), one equation of type (B) and all $10$ rank constraints. This formulation lead to a standard monomial basis of size $|\mathcal{B}|=22$. Despite only $3$ rank constraints would have been enough, numerical experiments revealed that using all $10$ equations leads to a more stable solver. Particularly, if only $3$ rank equations had been used, we would have obtained $|\mathcal{B}|=66$. Once the offsets and the parameters $\vec{H}$ and $\vec{b}$ have been determined, the receivers and the first $3$ transmitters can be determined with \eqref{eq:toa_upgrade}. The last transmitter $\vec{s}_4$ is finally computed using trilateration.


\section{Results}
\label{ch:results}
In this section, the experiments to quantitatively evaluate our solvers are presented. The solvers are benchmarked against synthetic data. The positions of the receivers and transmitters are sampled from a zero-mean normal distribution with standard deviation $10$. The time offsets are generated from a standard normal distribution. Our generated Matlab solvers lead to accurate solutions and run in $\sim\SI{200}{\milli\second}$ on an Intel i7-8565U processor, thus being suitable for near-real-time applications. 

In addition to the two proposed solvers, we also computed the standard monomial basis for configurations between the previously solved and the minimal ones, using the approach described in this paper. The results, depicted in Table \ref{tab:degree}, can be used to roughly assess the computational complexity of the unsolved cases, giving hints on their feasibility.
          

 \begin{table}[t]
    \centering
        \caption{Complexity estimate of different TDOA configurations. X: solved in \cite{kuang2013tdoa}.  -: unsolved u: underdetermined. *: reducible to an solvable configuration. Underlined the new solvers proposed in this paper.}
    \label{tab:degree}
    \begin{tabular}{c|cccc}
        \hline\hline
         $m/n$&3&4&5&6  \\\hline
         4&u&u&-&-\\
         5&u&154&64&X\\
         6&181&\underline{22}&*&*\\
         7&144&X&*&*\\
         8&116&*&*&*\\
         9&\underline{75}&*&*&*\\
    \end{tabular}
\end{table}     

\subsection{Clean data}
\begin{figure}[t]
     \centering
     \begin{subfigure}{0.49\linewidth}
        \includegraphics[width=\textwidth]{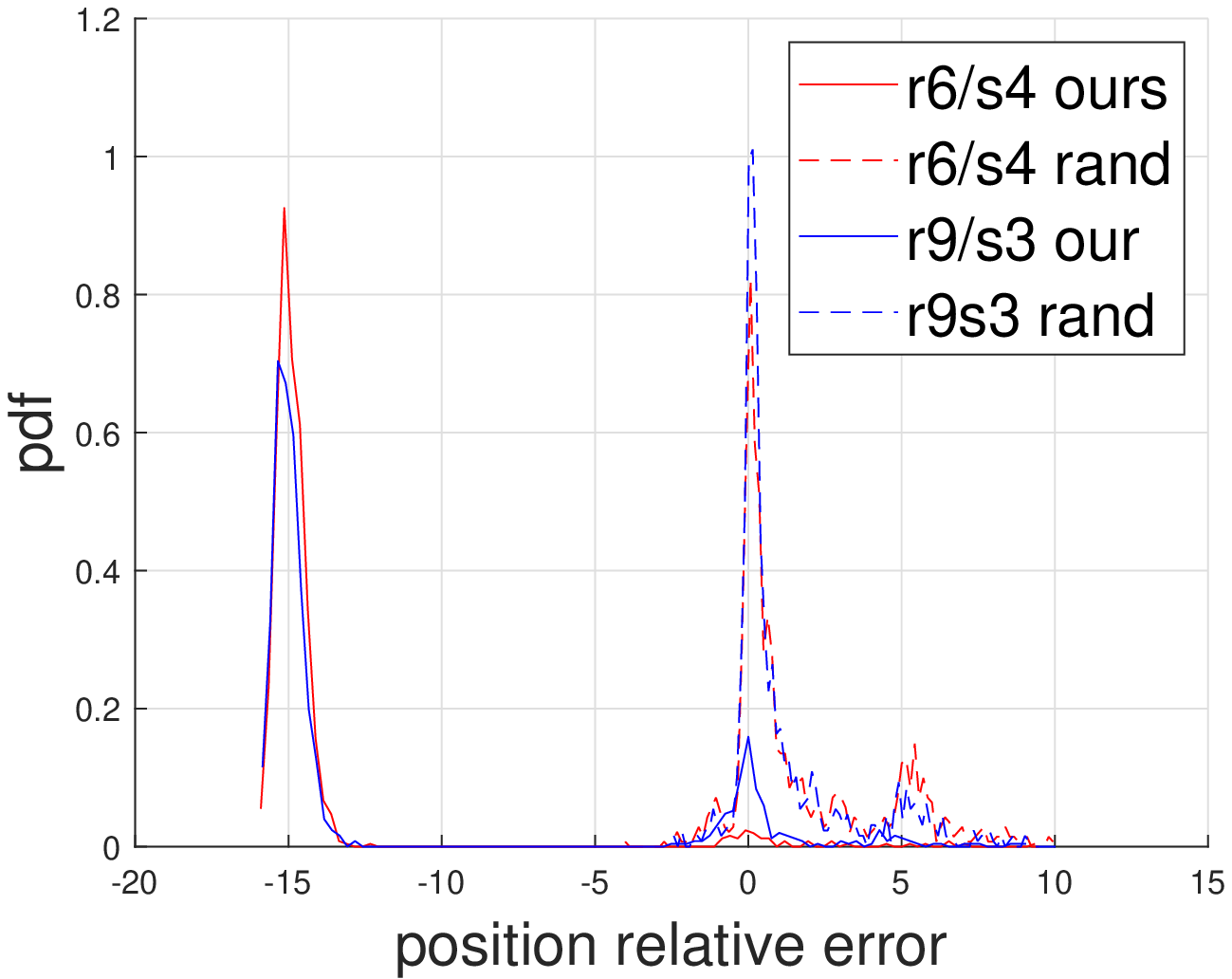}
     \end{subfigure}
     \begin{subfigure}{0.49\linewidth}
        \includegraphics[width=\textwidth]{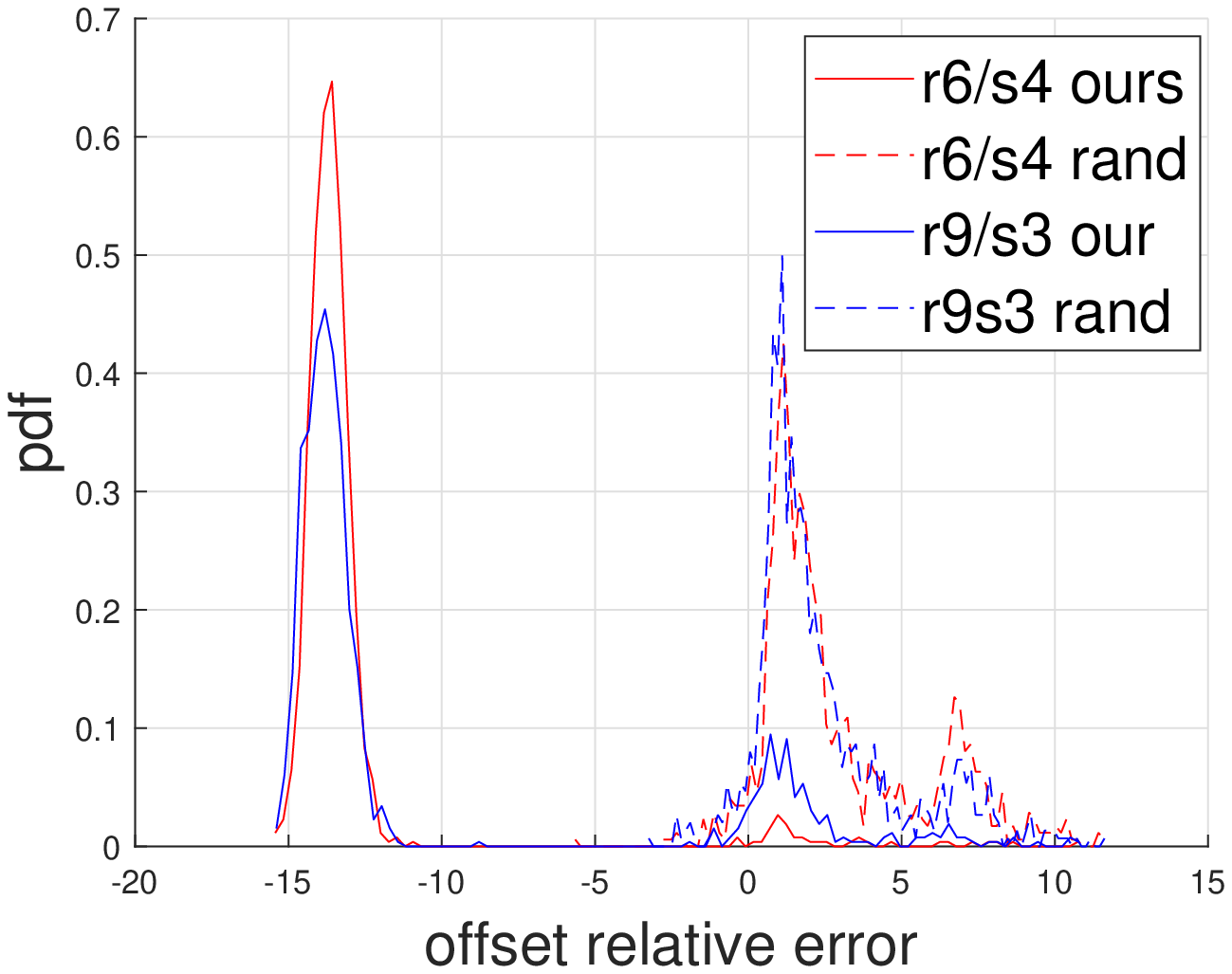}
     \end{subfigure}
     \caption{Relative error distribution for position and offsets when Levenberg-Marquardt is initialized with our solver (ours) and with a random initial value (rand).}
     \label{fig:histClean}
 \end{figure}

To evaluate our solvers, we randomly generate input data and solve the self-calibration problem using our proposed solvers. At this step, we consider noiseless data.  We also compare our solvers with the situation where the iterative algorithm is initialized with random positions and offsets, drawn from the same distributions of the ground truths. The relative error distributions, obtained running the solvers 1000 times with different random data, are shown in Fig. \ref{fig:histClean} and the median relative errors in Table \ref{tab:medianErr}. As can be noticed, even for clean data, a random initialization fails to converge. Our solver on the other side, produces accurate solutions, with the relative error being practically equal to the machine epsilon of the computer. Furthermore, using the solution of our solver as a starting point, the iterative algorithm converges already after very few iterations.

 \begin{table}[tbh]
     \centering
     \caption{Median relative errors in logaritmic scale of the proposed solvers.}
     \label{tab:medianErr}
     \begin{tabular}{c|c|c}
          Solver&Position error ($\log_{10}$)&Offset Error ($\log_{10}$) \\\hline\hline
          9r/3s rand& 0.31&1.63\\
          9r/3s our&-15&-14\\
          6r/4s rand&0.55&1.93\\
          6r/4s ours&-15&-14\\
     \end{tabular}
 \end{table}


 
 \subsection{Noisy data}
\begin{figure}[t!]
     \centering
     \begin{subfigure}{0.49\linewidth}
        \includegraphics[width=\textwidth]{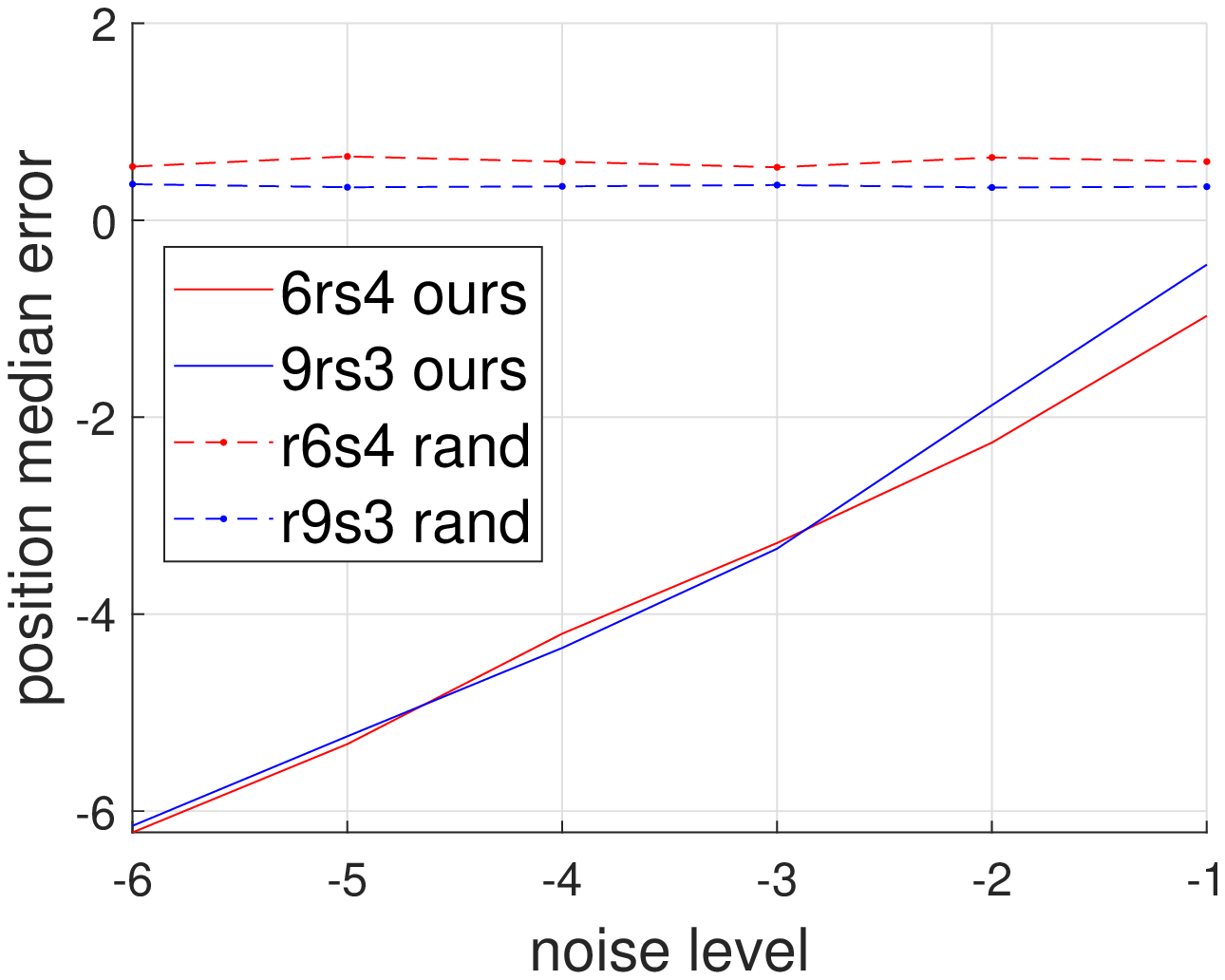}
     \end{subfigure}
     \begin{subfigure}{0.49\linewidth}  
        \includegraphics[width=\textwidth]{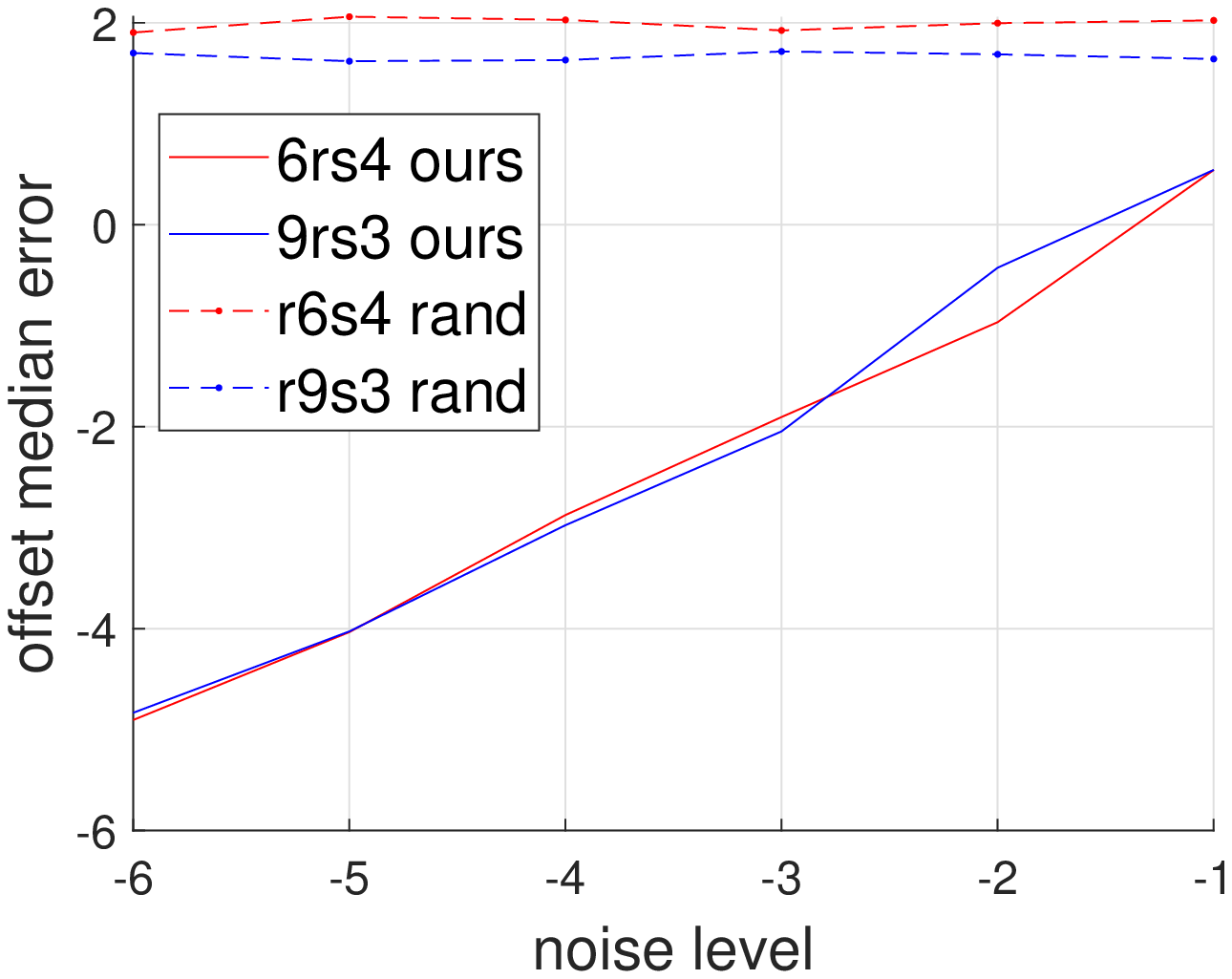}
     \end{subfigure}
     \caption{Median error as a function of $\sigma$ for position (left) and offsets (right).}
     \label{fig:noisy}
 \end{figure}

 We also investigate how our solvers perform with noisy data, adding zero-mean Gaussian noise with varying standard deviation $\sigma$ to the measurements $f_{ij}$. The results of the simulation are shown in Fig. \ref{fig:noisy}. As it was discussed before, TDOA problems are sensitive to the choice of the initial value and a poor initialization can cause convergence issues. While random initialization fails, using our solvers removes this problem, allowing fast and accurate convergence, even at higher noise levels.

\section{Conclusions}
\label{ch:conclusions}
In this paper we considered the sensor network self-calibration problem from 2D TDOA measurements. We proposed a novel algorithm which led to new robust and efficient polynomial solvers, which allowed to solve two previously unsolved configurations. We showed that the solutions obtained with our approach are stable both for clean and noisy data.

\section*{Acknowledgments}
This work was partially funded by the Academy of Finland project 327912 REPEAT.

\bibliographystyle{IEEEtranS}
\bibliography{ref}

\end{document}